\RequirePackage{fix-cm}
\documentclass[natbib,smallextended]{svjour3}       % onecolumn (second format)
\smartqed  % flush right qed marks, e.g. at end of proof
\usepackage{amsfonts,amsmath,amssymb,ulem,nicefrac,verbatim,color}
\usepackage{array,graphicx} %,longtable,ltcaption,lscape}

\usepackage{aps-bibstyle}  % use this style if you upload to .tex file only a part of Bibtex created bbl.
\def\numlimits{{23}}

\newcommand{\xraychap}{Saxton et al. (2020)}%{X-ray Chapter}
\newcommand{\gammachap}{Zauderer et al. (2020)}%{Gamma-ray Chapter}
\newcommand{\optchap}{van Velzen et al. (2020)}%{Optical Chapter}

\newcommand{\echochap}{van Velzen et al. (2019)}%{Echo Chapter}

\newcommand{\impostchap}{Zabludoff et al. (2020)}
\newcommand{\hostchap}{French et al. (2019)}
\newcommand{\ratechap}{Stone et al. (2019)}%{Rates Chapter}

\newcommand{\emischap}{Roth et al. (2020)}%{Emission Mechanisms Chapter}

 \journalname{ISSI Book on TDEs}
\usepackage{ref_macros} % svv added, needed for ADS style journal definitions
\begin{document}

\title{Radio Properties of Tidal Disruption Events}
%\subtitle{Subtitle}


\author{Kate D.~Alexander \and 
        Sjoert van Velzen         \and
        Assaf Horesh \and 
        B.~Ashley Zauderer
}


\institute{K.~D.~Alexander \at
              {Center for Interdisciplinary Exploration and Research in Astrophysics (CIERA) and Department of Physics and Astronomy, Northwestern University, Evanston, IL 60208} \\ 
{NASA Einstein Fellow} \\
              \email{kate.alexander@northwestern.edu}           %  \\
           \and
            S. van Velzen \at
              Center for Cosmology and Particle Physics, New York University, NY 10003
              \and
           A. Horesh \at
              The Racah Institute of Physics, The Hebrew University of Jerusalem, Jerusalem 91904, Israel
                         \and
           B.~A.~Zauderer \at
              {National Science Foundation, Division of Astronomical Sciences, 2415 Eisenhower Ave., Alexandria, VA 22314} \\ 
             {Dark Cosmology Centre, Niels Bohr Institute, University of Copenhagen, 
             Denmark} \\
              \email{bezauder@nsf.gov}
}

\date{Received: date / Accepted: date}

\maketitle

\begin{abstract}
Radio observations of tidal disruption events (TDEs) probe material ejected by the disruption of stars by supermassive black holes (SMBHs), uniquely tracing the formation and evolution of jets and outflows, revealing details of the disruption hydrodynamics, and illuminating the environments around previously-dormant SMBHs. To date, observations reveal a surprisingly diverse population. A small fraction of TDEs (at most a few percent) have been observed to produce radio-luminous mildly relativistic jets. The remainder of the population are radio quiet, producing {less luminous jets, non-relativistic} outflows or, possibly, no radio emission at all. Here, we review the radio observations that have been made of TDEs to date and discuss possible explanations for their properties, focusing on detected sources and, in particular, on the two best-studied events: Sw~J1644+57 and ASASSN-14li. We also discuss what we have learned about the host galaxies of TDEs from radio observations and review constraints on the rates of bright and faint radio outflows in TDEs. Upcoming X-ray, optical, {near-IR}, and radio surveys will greatly expand the sample of TDEs, and technological advances open the exciting possibility of discovering a sample of TDEs in the radio band unbiased by host galaxy extinction.

\end{abstract}


\newpage
 %\textit{We should emphasize here how improvements in radio sensitivity (EVLA) revolutionized TDE radio science, and point out that we are on the verge of a new revolution as we enter the era of large, sensitive radio surveys and instruments (ALMA, VLASS, MeerKAT, SKA, ngVLA, ASKAP, etc.}

\section{Introduction}

%\begin{itemize}
%\item Emphasize the unique diagnostic power of radio
%\item radio properties of TDEs vs SNe, AGN - {Ashley, Assaf, all.} \textit{Keep eye on overlap with imposters chapter.}

Radio observations have been used over the past several decades to detect and characterize outflows in extragalactic transient phenomena. These outflows can appear in different forms having a wide range of energies and velocities ranging from narrow ultra-relativistic jets (as in gamma-ray bursts; GRBs) to sub-relativistic spherical ejecta (as in core-collapse supernovae; SNe). Some famous examples in which radio observations provided key diagnostics of the type of outflows and helped reveal the nature of the transient phenomena in question include: the associated relativistic
outflow in the SN\,1998bw, the first SN found to be associated with a GRB \citep{gvv+99,kfw+98}; SN\,2009bb, a SN with a relativistic outflow that lacks $\gamma$-ray emission \citep{scp+10}; the discovery of both relativistic and sub-relativistic outflows in TDEs \citep{bgm+11,ltc+11,zbs+11,abg+16,vas+16} 
and the characterization of the
complex structure of the relativistic outflow in the first discovered neutron star merger GW170817 \citep{abf+17,amb+18,hcm+17,ksr+17,chl+18,dkm+18,max+18,mnh+18,mfd+18,mdg+18,rsi+18,tpr+18,gsp+19,hma+19}. 

It is now a common practice to perform radio observations of almost any
transient phenomena discovered, especially nearby ones, as the radio
emission is quite weak if the outflow is sub-relativistic. The radio
observations provide key diagnostics such as calorimetry, outflow velocity, magnetic field strength, and the density of the immediate environments surrounding the transient {(e.g. \citealt{chev82,chev98,fwk00,bnp13})}. In some cases
these crucial pieces of information are not accessible via other
wavelengths. Additionally, Very Long Baseline Interferometry (VLBI) radio observations can achieve higher resolution than available at any other wavelength, directly resolving the structure of jets and outflows for nearby events {(e.g. \citealt{tfb+04,tmp+05,ptg+07,rpc+16,mpe+18,mdg+18})}.

\begin{figure}
\centerline{\includegraphics[width=0.9\textwidth]{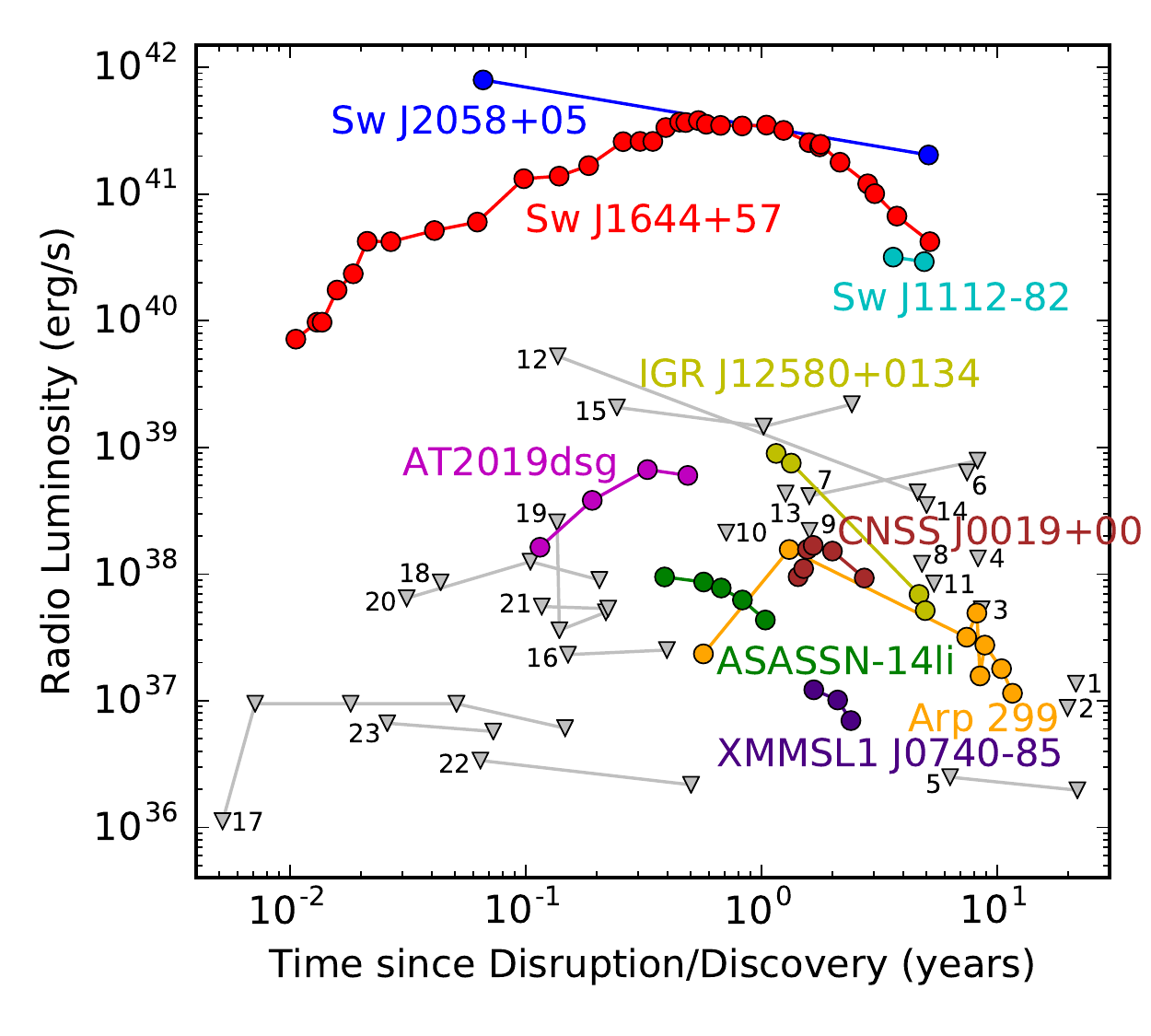}}
\vspace{-0.1in}
\caption{Literature TDE radio observations. To date, {nine} TDEs have published radio detections: Sw\,J1644+57, Sw\,J2058+05, Sw\,J1112-82, IGR J12580+0134, ASASSN-14li, XMMSL1 J0740-85, Arp 299-B AT1, {CNSS J0019+00, and AT2019dsg (colored circles; see Table \ref{tab:det} and references therein). Although most of the detected TDEs were observed at multiple frequencies, for simplicity we show only a single frequency for each event (8.4 GHz for Arp 299-B AT1 and AT2019dsg, 5 GHz for all others).} An additional \numlimits~events have published upper limits (gray triangles; {a key to the labels is given in the first column of Table \ref{tab:limits}). When a non-detected TDE was observed at multiple frequencies on the same date, we show only the most constraining limit.} All upper limits are $3\sigma$.}
\label{fig:radio}
\end{figure}

Radio observations of TDEs have revealed extremely diverse radio properties (Figure \ref{fig:radio}). This diversity and its implications are the subject of this review. We begin with an overview of the general synchrotron mechanism for producing radio emission (Section \ref{sec:sync}; {see also \emischap\,in this volume}). To illustrate the range of radio properties observed to date, we then present a detailed discussion of the two best-studied radio TDEs, Sw\,J1644+57 and ASASSN-14li, and briefly discuss the other TDEs with detected radio emission (Section \ref{sec:models}). We then discuss what we have learned about TDE host galaxies and TDEs' immediate environments from radio observations (Section \ref{sec:rho}). We finish with a discussion of occurrence rates of jets and outflows in TDEs and prospects for the future
(Section \ref{sec:rates}). %, including the implications of the most-constraining radio non-detections obtained to date.

\section{Radio Emission Basics}\label{sec:sync}

When a fast (either relativistic or sub-relativistic) outflow interacts with the circumstellar (interstellar) medium (CSM/ISM), it drives a shockwave into that medium. As the shockwave plows through the CSM/ISM, it accelerates free electrons to relativistic velocities and enhances magnetic fields. As a result, synchrotron emission ensues. This is the basic mechanism responsible for the radio emission in all types of transients mentioned above \citep[e.g.,][]{chev82,chev98,wpm+02,gs02,cf06,gm11,bnp13}, including TDEs. In most cases, synchrotron from the external shock between the expanding outflow and the surrounding medium dominates the observed emission, but sometimes synchrotron produced by internal shocks in the jet is also significant (e.g. in the compact cores of AGN jets, and possibly in some TDEs; see Section \ref{sec:jet} and \citealt{pv18}). {\emischap~in this volume provide additional theoretical discussion of synchrotron emission in the context of TDEs; here, we review the aspects most relevant for radio observations. }

The properties of the {observed synchrotron} emission depend on the physical properties of the system such as the CSM/ISM density, the velocity of the outflow driving the shockwave, the electron energy distribution, the total energy carried by the outflow, and other microphysical parameters such as the fractions of shockwave energy deposited in the electrons and the magnetic field, respectively. The resulting shape of the synchrotron radio spectrum is in general a multiple broken power-law, which can be fully described by its break frequencies and an overall normalization factor (e.g. \citealt{gs02}).\footnote{We make several simplifying assumptions, most importantly that the magnetic field is constant over the emitting region, and that all electrons are accelerated into a single power-law distribution of energies, {$N_e(E) \propto E^{-p}$ for $E > E_0$ ($E_0$ is typically taken to be $\sim$the electron rest mass energy).} This is typically the case for single impulsive events like GRBs and TDE jets, but is not always true. For example, the compact jet cores of AGN typically have flat spectra because we are observing a superposition of many different synchrotron components.} %{In particular, a synchrotron-emitting source at distance $d_L$ and redshift $z$ \citep{kn09}}.
{In particular, radio observations reveal the peak of the synchrotron emission spectrum, which for TDEs is typically where the emission transitions from optically thick ($F_{\nu}\propto\nu^{5/2}$ at low frequencies) to optically thin ($F_{\nu}\propto\nu^{-(p-1)/2}$ at higher frequencies). Relativistic bulk motion of the emitting region also affects the observed luminosity of the source, due to Doppler boosting effects. In particular, a source with a simple power-law spectrum $F_{\nu}\propto\nu^{-\alpha}$ that emits isotropically with flux density $F_0$ in its rest frame, moving relativistically with Lorentz factor $\Gamma\equiv1/\sqrt{1-\beta^2}$ at an angle $\theta$ to the observer's line of sight, will be observed to have a flux density $F$ enhanced by an amount $\delta^{2+\alpha} < F/F_0 < \delta^{3+\alpha}$ \citep{cr16}. Here, $\delta\equiv[\Gamma(1-\beta\cos\theta)]^{-1}$ is the Doppler factor of the source. For nearly on-axis jets, the Doppler boosting can be quite large, while the flux from off-axis jets will initially be relativistically beamed away from the observer, and the jet may not become observable until it decelerates and becomes non-relativistic. This explains why on-axis relativistic jets have higher observed peak luminosities than off-axis jets or sub-relativistic sources, and thus why on-axis jetted synchrotron sources (e.g.~GRBs, radio-loud TDEs) -- although rarer than non-relativistic sources (e.g.~SNe, some radio-quiet TDEs) -- can be detected to much higher redshifts.}

If the full synchrotron spectrum is observed and all of the break frequencies are measured (which typically requires coordinated radio, optical, and X-ray observations with dense temporal sampling), then all of the physical parameters listed above can be uniquely determined. If, as is often the case in TDEs, the optical and X-ray emission are dominated by other processes (see \emischap~in this volume), then radio and millimeter observations may provide the only constraints on the synchrotron spectrum and some break frequencies will not be observable, so some parameter degeneracies may persist. In such cases, we can make the simplifying assumption that the energy in the synchrotron-emitting outflow is in equipartition between the electrons and the magnetic field (e.g. {\citealt{sr77,chev98};} \citealt{bnp13}). By reducing the number of free parameters, this allows us to {estimate} the physical size of the emitting region, the kinetic energy of the outflow, and other physical properties (the outflow velocity, the ambient density, the average magnetic field strength, etc.) even if only part of the synchrotron spectrum is observed (preferably including the peak). The energy thus obtained is a lower bound on the total energy, which can be much larger if the source is not exactly in equipartition, while the size of the emitting region is more robust. Multi-frequency radio observations are preferred for this technique, to constrain the peak frequency and flux density of the radio emission and their temporal evolution. {Calculating} the size evolution of the emitting region allows us to infer when the outflow was launched {(assuming that the radio emission traces the leading edge of the outflow, as expected for external shock models).} This is an important constraint for modeling TDEs, for which the time of disruption may not be known precisely (e.g. \citealt{zbs+11,abg+16}). {For extremely nearby events, the size evolution of the outflow may also be measured directly using VLBI observations (e.g.~\citealt{mpe+18})}.

\section{Radio-detected TDEs: probes of accretion and outflow physics}\label{sec:models}

To date, several dozen TDEs have been observed in the radio, revealing a large diversity in their radio properties (Figure \ref{fig:radio}). In particular, a few percent of TDEs are radio loud, exhibiting luminous radio emission detectable for years post-disruption, while the rest are radio quiet, with detections or upper limits constraining their radio emission to be orders of magnitude fainter than the radio-loud events. {For the purpose of this review, we define a ``radio-loud TDE" to have a radio luminosity $\nu L_{\nu}>10^{40}$ erg s$^{-1}$ and a ``radio-quiet TDE" to have $\nu L_{\nu}<10^{40}$ erg s$^{-1}$. We list the nine TDEs with published radio detections in Table \ref{tab:det}, in order of decreasing peak radio luminosity.} While we generally assume that both radio-loud and radio-quiet TDEs are powered by the disruption of a star, the parameters of the disruption could be different. Radio-loud TDEs also exhibit bright high-energy emission and are discussed in more detail by \gammachap\ in this volume. Their unique properties are generally attributed to the fact that they launch very energetic relativistic jets {viewed on-axis}, while other TDEs {either launch jets viewed off-axis or do not launch energetic jets. (Off-axis jets will have a much fainter peak luminosity than on-axis jets because the radio emission is suppressed at early times by Doppler beaming, as discussed above.) Differences in parameters such as circumnuclear density, magnetic field strength or configuration, black hole spin, and disruption geometry may also contribute to the observed wide range of TDE radio luminosities (e.g.~\citealt{gm11,vkf11,vv13,kpc16,gmm+17,ysp+19}). The impact of host galaxy environment on TDE properties is discussed further in Section \ref{sec:rho} and in \hostchap~in this volume.} 

{In contrast,} radio-quiet TDEs are generally {(but not always)} discovered in the X-ray, optical, or UV (see \xraychap\ and \optchap\ in this volume). {Optical and X-ray-discovered} TDEs are also called thermal TDEs because their {optical, UV, and X-ray} emission generally can be well-modeled as a blackbody peaking in the UV or soft X-ray. Only a handful of thermal TDEs have radio detections, {but this may be partially an observational bias. Many TDEs received radio follow up that would have been too sparse or too shallow to reveal low-luminosity outflows similar to those seen in radio-detected sources (Figure \ref{fig:radio}, gray triangles), and about half of known TDEs have not been observed in the radio at all. While the current data cannot rule out the hypothesis that radio emission with a peak luminosity $\sim10^{38}$ erg s$^{-1}$ may be common in radio-quiet TDEs,} deep upper limits for several recent events (e.g. \citealt{blag17,vv18,srk+19}) hint that there may be additional diversity within this class. We have compiled a list of all published radio non-detections of TDEs available in the literature, presented in Table \ref{tab:limits}. Below, we discuss the physical insights derived from radio observations of the two best-studied radio-detected TDEs to date: the radio-loud Sw\,J1644+57 and the radio-quiet ASASSN-14li. We also briefly summarize the properties of other TDEs detected in the radio.

\subsection{Sw\,J1644+57: A relativistic jet}\label{sec:1644}

Sw\,J1644+57 was discovered by the Neil Gehrels Swift Observatory on 28 March 2011 as an unusual $\gamma$-ray and X-ray transient coincident with the nucleus of a galaxy at $z=0.354$ \citep{bgm+11,bkg+11,ltc+11,zbs+11}. Associated radio emission was discovered within 24 hours, initiating an extensive radio observing campaign spanning $1-345$ GHz \citep{zbs+11,bgm+11,ltc+11,bzp+12,wvl+12,z13,ypv+16,ebz+18}\footnote{Sw\,J1644+57 was also observed at 149 MHz by LOFAR, but the resulting non-detection is not constraining \citep{cws+14}.}. The full radio dataset includes observations extending to $\approx2000$ days post-discovery and the source continues to be detected in multi-frequency observations. Here, we focus on the radio properties of Sw J1644+57; for more discussion of this source and other TDEs with high-energy emission, see \gammachap\ in this volume.

The radio observations of Sw J1644+57 capture the peak of the synchrotron emission. The peak frequency, $\nu_p$, decreases with time, while the peak flux density $F_{\nu,p}$ remains approximately constant (within a factor $\sim3$) for $\sim400$ days and then decreases \citep{ebz+18}. This behavior is consistent with synchrotron emission from an initially mildly relativistic ($\Gamma\approx3$), collimated jet that expands and slowly decelerates \citep{zbs+11,bzp+12,z13,ebz+18}. The jet launch date inferred from the radio observations is consistent with the onset of the high-energy emission \citep{zbs+11}. For equipartition models and other models where the energy carried by electrons and the energy in magnetic fields remain constant fractions of the total energy, the total energy must increase by an order of magnitude from $30-250$ days. This could be explained by structured ejecta, in which slower-moving material catches up to the head of the jet as it decelerates and begins contributing to the emission at later times (e.g. \citealt{bzp+12,mgm+15}). An alternate possibility is that the energy distribution varies in time; e.g. most energy is initially carried by the magnetic field, and at later times shifts to being carried by the electrons \citep{bp13}. In this case, it is possible for the total energy in the jet to remain constant. However, more recent work sees no evidence for such time variations \citep{ebz+18}.

The electrons producing the radio emission in Sw J1644+57 are in the slow cooling regime, but at higher frequencies a cooling break in the synchrotron spectrum is expected (e.g. \citealt{gs02}). The location of this break provides an additional constraint on the synchrotron model, breaking parameter degeneracies and removing the need to assume equipartition. The optical data alone cannot precisely constrain the position of the cooling break, as they are affected by a large but uncertain amount of host extinction \citep{ltb+16}. Until 500 days, the X-rays are dominated by emission from the base of the jet \citep{zbs+11,bzp+12,z13}, rather than by the same synchrotron component producing the radio emission. At 500 days, the X-ray luminosity drops precipitously as the jet shuts off, and the residual X-rays are consistent with an extension of the radio emission spectrum \citep{z13}. The post-500 days X-ray data, together with the optical observations, allow constraints on both the cooling break frequency and the amount of host extinction \citep{ebz+18}. This reveals a magnetic field lower than the equipartition value, and correspondingly increases the total energy in the jet by a factor of $3$ to $E\sim4\times10^{51}$ erg \citep{ebz+18}.

\subsection{ASASSN-14\lowercase{li}: A thermal TDE with radio emission}\label{sec:14li}

ASASSN-14li was discovered by the All Sky Automated Survey for SuperNovae (ASAS-SN) on 22 November 2014, coincident with the center of the $z=0.0206$ galaxy PGC043234 (luminosity distance $d_L \approx 90$ Mpc). It was subsequently detected across the electromagnetic spectrum (\citealt{mkm+15,hkp+16,vas+16,abg+16,jiang16}; {\citealt{rpc+16}}; \citealt{bright18}). Unlike many TDEs, it exhibited both optical and X-ray variability, along with declining radio emission and an infrared echo. ASASSN-14li has the best-studied radio emission of all thermal TDEs and has become a poster child for this class. Below, we summarize several different models that have been proposed to explain its radio emission, which is $\sim10^3$ times less luminous than Sw\,J1644+57's. 

\subsubsection{Non-Relativistic Wind}\label{sec:wind}

\cite{abg+16} modeled ASASSN-14li's radio emission as synchrotron radiation arising from the external shock between outflowing material from the TDE and the surrounding circumnuclear medium. In this model, the peak of the radio spectral energy distribution (SED) corresponds to the synchrotron self absorption frequency, which depends on the density and the properties of the outflow. Assuming that the energy contained in the outflow is always in equipartition between the electrons and the magnetic field (and that emission from a single region dominates the radio SED at all frequencies) allows the energy and radius of the emitting region to be computed independently in each multi-frequency radio epoch (e.g. \citealt{bnp13}). Thus, no dynamical assumptions are required and the derived physical properties of the emitting region don't depend on any specific launching mechanism for the outflow.  

In the case of ASASSN-14li, multi-frequency radio observations with the Karl G. Jansky Very Large Array (VLA) {were found to be consistent with} an outflow with a {constant equipartition} energy of $\sim10^{48}$~erg expanding at a constant velocity $0.04-0.12c$ \citep{abg+16}. (There is a mild degeneracy between the energy/velocity and the assumed opening angle of the outflow, with a spherical outflow moving slower than a conical one.) This method also revealed a steep circumnuclear density profile proportional to $r^{-2.5}$ (Section \ref{sec:rho}). These properties are consistent with expectations for a wind produced during a phase of super-Eddington accretion \citep{sq09}. \cite{abg+16} showed that the launch time of ASASSN-14li's outflow roughly coincided with the time when the accretion rate first exceeded Eddington (independently estimated from the optical, UV, and X-ray light curves), providing additional support for this picture. In this scenario, only TDEs in which the peak accretion rate exceeds the Eddington limit would produce radio emission. Alternately, recent simulations suggest that a similar outflow could also be launched by the self-intersection of the bound debris stream, expected due to relativistic apsidal precession \citep{lb19}. Such collision-induced outflows could produce radio emission with a wide range of peak luminosities and timescales.

\subsubsection{Unbound Tidal Debris Stream}\label{sec:unbound}

Approximately half of the stellar debris from a TDE will ultimately accrete onto the SMBH, while the rest is unbound \citep{rees88}. A second possibility is that ASASSN-14li's radio emission is produced by the interaction between the unbound debris and the circumnuclear medium. In this case, synchrotron radiation is emitted in the bow shock that forms along the leading edge of the debris stream, and the emitting region is expected to expand at a velocity $\sim0.2c$ \citep{kpc16}. \cite{kpc16} performed a similar equipartition analysis to \cite{abg+16} and reached similar conclusions about the velocity and energetics of the radio-emitting material in ASASSN-14li, roughly consistent with this picture. Only a tiny fraction of the unbound debris is required to produce the observed radio luminosity, consistent with a model in which the bulk of the unbound debris has not yet decelerated on the timescale of the observations.

An important prediction of this model is that all TDEs should generate radio emission from this process, with no dependence on the peak accretion rate. However, the brightness of the emission will depend on both the ambient density and the cross-section of the debris stream (which may be quite small for some observer viewing angles). Some theoretical studies suggest that in most disruptions the debris stream should remain self-gravitating until large radii (e.g. \citealt{gmc16,scs+19}); in this case, the stream may be too narrow to produce detectable radio emission. However, for disrupted stars on deeply plunging orbits, the unbound debris fan may be more dispersed \citep{sq09,ysp+19} and more luminous emission similar to that seen in ASASSN-14li may be possible. As only a small amount of mass is required to produce the observed radio emission in ASASSN-14li \citep{kpc16,ysp+19}, only a small fraction of the ejecta need not be self-gravitating. Recent work by Piran et al. (in prep) suggests that in most TDEs, radio emission from the bow shock generated by the fastest unbound debris should be marginally detectable, with a peak flux density of a few $\mu$Jy for a TDE at the distance of ASASSN-14li ($\sim90$ Mpc).

\subsubsection{Jet}\label{sec:jet}

A third possibility to explain the radio flare of ASASSN-14li is emission from a (sub-)relativistic jet. Such jets can have two modes of emission. External emission is powered by a forward shock that the jet drives into the surrounding medium (in AGN jets these shocks power the so-called hotspots). The external mode of emission is akin to the wind and unbound-debris models discussed above. However jets can also produce synchrotron emission without interaction with a surrounding medium. This internal emission mechanism is powered by shocks inside the jet (in AGN jets, internal emission is responsible for the compact, flat-spectrum radio cores). 

The first model that was proposed for radio emission of ASASSN-14li was external (i.e., shock-driven) emission from a relativistic jet \citep{vas+16}. This model could explain the 16~GHz light curve, but the multi-frequency radio observations \citep{abg+16} that appeared later challenged this interpretation. {\cite{rpc+16} also suggested the possibility of a relativistic jet in ASASSN-14li, based on VLBI observations that showed two separate components of radio emission separated by $\sim2$ pc (implying superluminal motion from a nearly on-axis relativistic jet with $\Gamma \gtrsim 7$). However, jets with similarly high Lorentz factors typically have much higher brightness temperatures, so they conclude that the second emission component is more likely unrelated to the TDE (Section \ref{sec:vlbi}).}

An improvement to the jet model for ASASSN-14li is presented in \cite{pv18}, who updated the original internal emission mechanism for TDE jets \citep*{vkf11} to include adiabatic cooling. In this model the jet consists of a super-position of synchrotron-emitting regions in a conical geometry, each with their own equipartition magnetic field. Applying this model to the observed synchrotron SEDs yield a measurement of the scaling of the equipartition magnetic field with the radius in the jet, which was found to be $B\propto R^{-1.02\pm0.05}$. This is remarkably close to the predicted power-law index of $-1$ for a freely expanding jet that is powered by an internal emission
mechanism \citep{bk79,fb95}.  

The strongest evidence against an external emission mechanism for the radio properties of ASASSN-14li is the detection of a cross-correlation between the X-ray and radio light curves \citep{pv18}; the radio lags the X-rays by about 12 days (see \echochap\ in this volume for details of the cross-correlation). Establishing this correlation requires a coupling between the X-ray emitting disk and the source of the radio photons. This coupling is not expected in any of the proposed external emission mechanisms (disk-wind, streams shocks, or jet hotspots) because these are all impulsive: the radio emission is governed by the evolution of the shock, which decouples from the accretion disk. However in an internal jet model, this coupling is naturally expected (and is observed for jets from active black holes; \citealt{fb95,m+02}). The observed 12 day lag is consistent with the time that is required for the 16~GHz radio emission to become optically thin to synchrotron self absorption.  

\begin{center}
\begin{table}
\caption{{TDEs with Published Radio Detections.}}
\label{tab:det} 
\begin{tabular}{lcccll}
%\hline \noalign{}
\hline\noalign{}
\hline\noalign{\smallskip}
{Name} & ${z}$ & {Discovery} & {Peak radio} & {Proposed origin(s)} & {Reference(s)} \\
 & & {method} & {luminosity} & {of radio emission} & {for radio data} \\
 & & & {(erg s$^{-1}$)} & {(see text)} & \\
 \hline\noalign{}
\hline\noalign{\smallskip}
{Sw J2058+05} & {1.1853} & {$\gamma$-rays} & ${8\times10^{41}}$ & {on-axis initially} & {\cite{ckh+12};}\\
 & & & & {relativistic jet} & {\cite{pcl+15};} \\
  & & & & & {\cite{bls+17}} \\
\hline\noalign{\smallskip}
{Sw J1644+57} & {0.354} & {$\gamma$-rays} & ${4\times10^{41}}$ & {on-axis initially} & {\cite{zbs+11};}\\
 & & & & {relativistic jet} & {\cite{bzp+12};}\\
 & & & & & {\cite{wvl+12};}\\
 & & & & & {\cite{z13};}\\
 & & & & & {\cite{ypv+16};}\\
 & & & & & {\cite{ebz+18}}\\
\hline\noalign{\smallskip}
{Sw J1112$-$82} & {0.89} & {$\gamma$-rays} & ${3\times10^{40}}$ & {on-axis initially} & {\cite{bls+17}}\\
 & & & & {relativistic jet} & \\
\hline\noalign{}
\hline\noalign{\smallskip}
{IGR 12580+0134*} & {0.00411} & {X-rays} & ${9\times10^{38}}$ & {off-axis initially} & {\cite{ihk+15};}\\
 & & & & {relativistic jet} & {\cite{ywl+16};} \\
  & & & & & {\cite{pmw+17}} \\
 \hline\noalign{\smallskip}
{AT2019dsg} & {0.051} & {optical} & ${7\times10^{38}}$ & {outflow (collimation} & {\cite{Stein20}}\\
 & & & & {and max velocity} & \\
 & & & & {uncertain)} & \\
\hline\noalign{\smallskip}
{Arp 299-B AT1*} & {0.010411} & {near-IR} & ${2\times10^{38}}$ & {off-axis jet} & {\cite{mpe+18}}\\
\hline\noalign{\smallskip}
{CNSS J0019+00*} & {0.018} & {radio} & ${2\times10^{38}}$ & {non-relativistic} & {\cite{amh+19}}\\
 & & & & {outflow} & \\
 \hline\noalign{\smallskip}
{ASASSN-14li*} & {0.0206} & {optical} & ${9\times10^{37}}$ & {non-relativistic} & {\cite{abg+16};}\\
 & & & & {outflow, internal} & {\cite{vas+16};}\\
 & & & & {shocks in on-axis jet,} & {\cite{rpc+16};}\\
 & & & & {or unbound debris} & {\cite{bright18}}\\
\hline\noalign{\smallskip}
{XMMSL1 J0740-85} & {0.0173} & {X-rays} & ${1\times10^{37}}$ & {outflow (collimation} & {\cite{srk+17};}\\
 & & & & {and max velocity} & {\cite{awb+17}} \\
 & & & & {uncertain)} & \\ 
\hline\noalign{}
\hline\noalign{\smallskip}
\end{tabular}
{*host galaxy exhibited weak to significant AGN activity prior to the onset of the TDE.}
\end{table}
\end{center}

\subsection{Other Detected Sources}\label{sec:other}

At the time of writing, seven other TDEs with radio detections have been reported in the literature (Table \ref{tab:det}). Two of these sources, Sw J2058+05 and Sw J1112-82, were discovered in archival searches of Swift data and have similar $\gamma$-ray, X-ray, and radio properties to Sw J1644+57 \citep{ckh+12,pcl+15,bls+15,bls+17}. Like Sw J1644+57, they likely launched energetic relativistic jets viewed on-axis. For a more complete discussion of these sources, see \gammachap.

As jets in the most energetic TDEs are thought to be narrowly collimated, we expect that most jetted TDEs will be viewed off-axis. In this case, the radio emission is suppressed at early times due to relativistic beaming effects, but will appear similar to on-axis jetted events after the jet decelerates  \citep{gpk+02,vv13}. However, by this point, the radio emission evolves very slowly -- so it may be difficult to identify the transient nature of off-axis TDEs \citep{ebz+18}. To date, two additional radio TDEs, IGR J12580+0134 and Arp 299-B AT1, have been proposed to exhibit off-axis jets \citep{lei16,mpe+18}. IGR J12580+0134's radio emission was discovered serendipitously in a survey of nearby galaxies \citep{ihk+15}, and was connected to a hard X-ray flare interpreted as a TDE that occurred in the same galaxy one year prior \citep{nw13}. Its peak radio luminosity is intermediate between Sw J1644+57's and ASASSN-14li's (Figure \ref{fig:radio}). This event occurred in a known AGN, so it is possible that the radio emission is associated with another type of AGN flaring event -- nevertheless, late-time radio follow up observations show that the radio flux density has decreased by a factor $\gtrsim10$ over five years \citep{pmw+17}, which is extreme for typical AGN flares \citep{hnt+08,nht+09}. 

In contrast, Arp 299-B AT1 was discovered as a bright flare in the near-IR, and {was luminous and long-lasting in both the IR and radio \citep{mpe+18}. It} exhibits no variability at {shorter} wavelengths due to extremely high host {galaxy} extinction \citep{mpe+18}. Its host, Arp 299, consists of two merging galaxies, potentially resulting in a temporarily enhanced TDE rate; the nucleus in which the flare occurred is also a {Compton-thick} AGN with an {almost} edge-on torus. It has been observed extensively in the radio with VLBI, directly imaging a resolved jet in a TDE for the first time \citep{mpe+18}. These observations initially show only a single emerging jet with no corresponding counterjet, allowing a constraint on the viewing angle ($25-35^{\circ}$). This implies that the TDE accretion disk is not aligned with the pre-existing AGN torus. It is possible that many TDEs occur in galaxy nuclei with similarly high levels of extinction, and are currently being missed by the optical and X-ray surveys that have yielded most TDE discoveries to date. 

{The remaining three radio-quiet TDEs also likely launched outflows, but these outflows may be quasi-spherical rather than collimated jets.} XMMSL1 J0740-85 was discovered during the XMM Newton Slew Survey \citep{saxton08} on 2014 April 1 as a bright X-ray transient \citep{srk+17}. Its X-ray spectrum contained both thermal and non-thermal components, placing it in between the jetted Swift events and TDEs not detected in $\gamma$-rays. XMMSL1 J0740-85 exhibited little to no variability in the optical and modest variability in the UV \citep{srk+17}. At late times ($\gtrsim600$ days after discovery), weak, fading radio emission was observed \citep{awb+17}. The origin of this radio emission is ambiguous: it is consistent with either a decelerated jet or an always non-relativistic outflow. In either model, the energy contained in the outflowing material is much less than the energy of Sw\,J1644+57's jet.

Further deep radio observations of TDEs are needed to determine whether weak radio emission like that seen in XMMSL1 J0740-85 and ASASSN-14li is common (Section \ref{sec:rates}), and to determine the physical conditions that lead to jet formation in a small subset of TDEs.  {Such efforts are already proving fruitful. For example, the most recently reported optical TDE with a radio detection, AT2019dsg \citep{Stein20}, has a similar luminosity to ASASSN-14li, but unlike ASASSN-14li it shows a linear increase of the equipartition energy with time, possibly suggesting that the outflow or jet in this source receives a constant injection of power at its origin. }

Future radio facilities will have the capabilities to carry out blind searches for radio transients, and may discover many events {that would have been missed by optical and X-ray surveys. Radio emission is not affected by extinction, so a uniform sample of TDEs \textit{discovered} in the radio will provide an independent estimate of the TDE rate} (Section \ref{sec:blind}). {The final radio-detected TDE, CNSS J0019+00, was discovered in a blind search for radio transients in the SDSS Stripe 82 region, illustrating the promise of this technique \citep{mhb+16,amh+19}. This TDE was in a Seyfert 2 galaxy and was localized to within 1 pc of the galaxy nucleus by VLBI observations. CNSS J0019+00's luminosity and duration make it more similar to other radio-quiet TDEs than to supernovae or AGN flares (although a reactivated AGN cannot be completely ruled out; \citealt{amh+19}). Similar to ASASSN-14li and XMMSL1 J0740-85, CNSS J0019+00's radio emission can be well-modeled as a sub-relativistic spherical outflow with an energy $\sim10^{49}$ erg, expanding at a velocity $\sim0.05c$ \citep{amh+19}.}

\section{Host properties and constraints on nuclear environments from radio observations}\label{sec:rho}

Radio observations of TDEs also reveal useful information about the properties of their host galaxies. The immediate vicinity of the SMBH can be probed via the interaction between outflowing debris from the TDE and the circumnuclear medium, while VLBI observations may reveal structures in the circumnuclear environment at larger scales. Processes such as star formation and AGN activity may also produce radio emission in the vicinity of the TDE; these additional sources of emission must be properly accounted for when interpreting radio observations carried out during the flare. Due to the long cooling time, relic lobes of AGN activity prior to the TDE can generate (near) nuclear radio emission that lingers for $\sim$Gyr. Additionally, AGN can exhibit radio variability on timescales ranging from hours to decades, including flares with typical timescales of a few years \citep{hnt+08,nht+09}. The amplitude of such flares is typically much smaller than the level of variability seen during radio-detected TDEs, but more observations are needed to properly locate radio TDEs within the broader framework of SMBH accretion. 

Surveying large areas of the radio sky at high resolution is time-consuming and data-intensive; until recently, only shallow extragalactic source catalogs existed for the vast majority of the radio sky. Thus, meaningful constraints on the pre-flare radio emission from TDE hosts has so far only been possible for the nearest events. This situation has started to change with the advent of the VLA Sky Survey (VLASS), which will image the entire sky north of declination $-40^{\circ}$ with a combined limiting rms of 69 $\mu$Jy, {and with the deployment of ASKAP and MeerKAT in the southern hemisphere, building towards the extremely sensitive Square Kilometer Array (SKA)}. In this section, we summarize the progress that has been made so far using radio observations to study the nuclear environments in which TDEs occur.

\subsection{Density Constraints}

When radio emission is produced in an external shock between a transient outflow and the surrounding medium, then we can also use radio observations of TDEs to infer the ambient density surrounding the supermassive black hole at parsec or even sub-parsec scales (e.g. \citealt{abg+16,kpc16,ebz+18}; {\citealt{mpe+18,amh+19,Stein20}}). 
As the outflow expands, independent measurements of the density at different epochs can be used to construct a density profile of the circumnuclear gas. (The dust distribution at comparable distances can be indirectly probed via the detection of infrared dust echoes; see \echochap { in this volume and references therein}.) % \citealt{lu16,vmk+16,jiang16,dwj+16}.) 
Such constraints are extremely valuable, as these scales are not directly resolvable at any wavelength with current facilities at the distance of most TDE hosts.

\begin{figure}
\centerline{\includegraphics[width=0.9\textwidth]{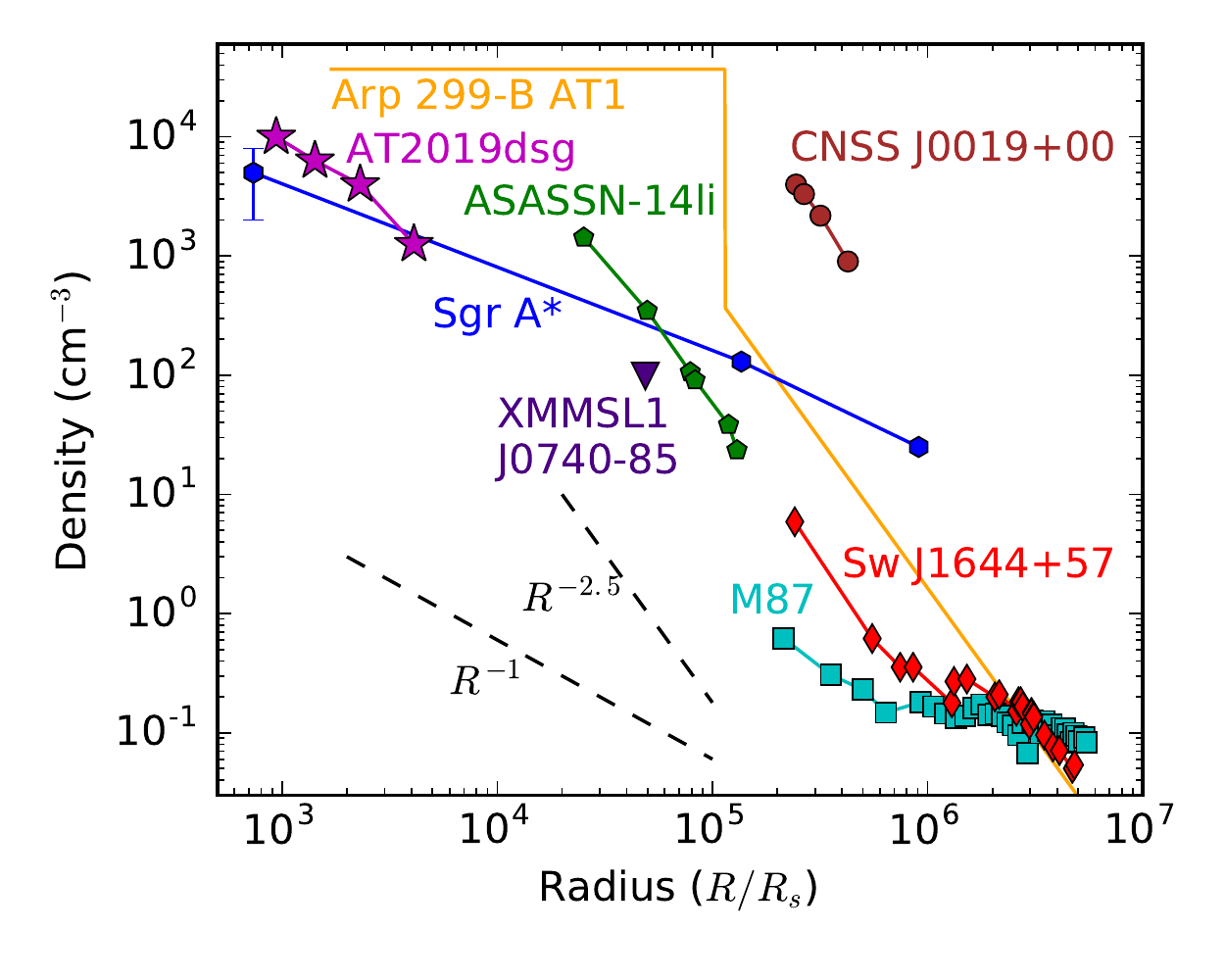}}
\vspace{-0.2in}
\caption{The circumnuclear density profiles of TDE host galaxies inferred from their radio observations, with the distance from each SMBH given in units of its Schwarzchild radius. The normalization of the profiles for ASASSN-14li \citep{abg+16}, XMMSL1 J0740-85 \citep{awb+17}, {and AT2019dsg \citep{Stein20}} depends on the geometry of the outflows, {which is not well constrained by the data; here, we assume} spherical outflows. The density profile for Sw J1644+57 is taken from the analysis by \cite{ebz+18}, {the profile for Arp 299-B AT1 is taken from \cite{mpe+18}, and the profile for CNSS J0019+00 is taken from \cite{amh+19}}. Shown for comparison are measurements of the density profile around two other well-studied SMBHs: the nearby and more massive M87 \citep{rfm+15}, and Sagittarius A* \citep{bmm+03,gpw+19}. Sgr A* has a much shallower density profile ($\rho \propto r^{-1}$) than ASASSN-14li's SMBH ($\rho \propto r^{-2.5}$) at comparable distance scales. {We also show sample $r^{-1}$ and $r^{-2.5}$ profiles to guide the eye (black dashed lines).}}
\label{fig:density}
\end{figure}

The shape of the ambient circumnuclear density profile directly probes the history of accretion onto the SMBH in the TDE host galaxy and the circumnuclear environment. Spherically symmetric accretion results in a Bondi profile, $\rho \propto r^{-3/2}$ \citep{bondi52}, while adding the effects of mass loss from massive stars produces a shallower profile (e.g. \citealt{q04,gmm+17}). A recent compilation of density measurements for Sagittarius A* reveals a density profile $\rho\propto r^{-1}$ over four orders of magnitude in distance\footnote{{We note that the available observations only sparsely sample Sgr A*'s density profile, so in principle deviations from $r^{-1}$ over a narrower range of radii could be present.}}, consistent with this picture (\citealt{gpw+19} and references therein). We compare these results with the density measurements inferred for TDE host galaxies in Figure \ref{fig:density}. {AT2019dsg's density profile is similar to Sgr A*'s \citep{Stein20}, while} Sw J1644+57's host galaxy exhibits a Bondi-like profile at $\gtrsim0.7$ pc \citep{ebz+18}. In contrast, ASASSN-14li's {and CNSS J0019+00's} host galaxies exhibit a steeper profile, $\rho\propto r^{-2.5}$ (\citealt{abg+16,kpc16}; {\citealt{amh+19}}).\footnote{We note that this result is only valid for models in which the radio emission arises from an external (forward) shock (Sections \ref{sec:wind} and \ref{sec:unbound}). In the model presented by \cite{pv18} (Section \ref{sec:jet}), the radio emission arises internal to the jet and thus does not provide any information about the density profile of the external medium. However, because the data prefer a freely-expanding jet, \cite{pv18} infer that the external density must be low.} {Arp 299-B AT1's radio emission can be best fit with a dense torus extending to $6.75\times10^{17}$ cm, with a similarly steep $r^{-2.5}$ density profile at larger radii \citep{mpe+18}}. Such a steep profile is not expected for spherically symmetric accretion, but appears in some models of super-Eddington accretion flows (e.g. ZEBRAs; \citealt{cb14}). This may indicate that the circumnuclear environments of ASASSN-14li's{, Arp 299-B AT1's, and CNSS J0019+00's} host galaxies have been shaped by previous episodes of accretion onto their SMBHs. {This is consistent with multi-wavelength evidence of at least a low level of AGN activity in these galaxies that pre-dates the TDE outburst \citep{abg+16,rpc+16,mpe+18,amh+19}.}

\subsection{Radio evidence for AGN activity and complex nuclear environments}\label{sec:vlbi}

{Interestingly, of the six published radio-quiet TDEs with detected low-luminosity radio emission, only two (XMMSL1 J0740-85 and AT2019dsg) occurred in host galaxies that showed no signs of recent or ongoing AGN activity (Table \ref{tab:det}). This may indicate that AGN produce environmental conditions particularly favorable for the production of detectable radio emission in TDEs, but it also raises the question of possible contamination. While this is a valid concern (as noted in Section \ref{sec:other}, AGN flares cannot be completely ruled out as an explanation for two or three of the radio-quiet TDEs discussed here), the past two decades of increasingly detailed observations have also allowed us to build up a set of multi-wavelength characteristics that differentiate AGN variability from true TDEs. The multi-wavelength properties that distinguish TDEs from ``imposters" such as nuclear SNe or AGN flares are discussed in detail in \impostchap~in this volume. For some events whose optical and X-ray properties are ambiguous, the radio properties of the transient and its host galaxy may help to clarify its nature, as AGN variability is typically much less extreme at radio wavelengths compared to other parts of the electromagnetic spectrum (e.g.~\citealt{hnt+08,nht+09}).}

Pre-flare radio observations of a TDE's host galaxy provide important context for interpreting radio observations collected during the TDE {\citep[e.g.,][]{abg+16,vas+16,rpc+16,mpe+18,amh+19}}. For example, ASASSN-14li's host galaxy was detected in two archival radio surveys $15-20$ years prior to the discovery of the TDE. Optical spectra of the host galaxy reveal a level of ongoing star formation an order of magnitude too low to explain the archival radio emission, so the best explanation for the archival detections is that ASASSN-14li's host galaxy hosts a weak AGN \citep{abg+16,vas+16}. Long-term monitoring of ASASSN-14li's radio emission revealed a plateau starting $\sim1$ year after discovery, indicating that the transient component of the emission has faded, and we are now viewing only the quiescent component \citep{bright18}.

The spectral index of the baseline radio emission in the host galaxy of ASASSN-14li is steep ($\alpha\approx -1$), suggesting that the emission is due to a ``relic" population of electrons that was accelerated a up to few Gyr prior to the TDE. Integral field spectroscopy supports this evidence for recent, but not on-going, AGN activity in the host galaxy of ASASSN-14li \citep{pka+16}. VLASS observations of new TDEs will help determine whether similar low levels of AGN activity are common in TDE host galaxies. 

A more detailed picture of the nuclear environment may also be obtained via extremely high resolution VLBI observations. VLBI observations at early times can help pinpoint the nuclear position of the transient, confirming or ruling out a TDE origin. This is especially useful for more distant events, like Sw J1644+57 \citep{zbs+11}. Observations at later times can also constrain the velocity and orientation of the jet relative to the observer's line of sight {(as \citealt{ypv+16} do for Sw J1644+57)}, and can even directly resolve the jet for nearby events ({as \citealt{mpe+18} do for Arp 299-B AT1)}.

VLBI observations of nearby events additionally reveal that many TDEs occur in complex nuclear environments. In two events, ASASSN-14li and IGR12580, {VLBI observations taken after the discovery of each transient} revealed multiple emission components in the nuclear vicinity, some of which were at least marginally resolved \citep{rpc+16,pmw+17}. In both cases, these secondary components would require superluminal motion if associated with the TDE; more likely they are related to previous episodes of accretion activity. In the case of ASASSN-14li, it is also possible that the two observed radio components may correspond to the two components of a binary black hole \citep{rpc+16}. {Additionally, \cite{mpe+18} observed multiple compact radio sources in the nuclear vicinity (within a few parsecs) of Arp 299-B AT1 in pre-TDE VLBI imaging. This is not surprising, given that its host galaxy both contains an AGN and has an exceptionally high core-collapse supernova rate.}

\section{Upper Limits and Occurrence Rates}\label{sec:rates}

\subsection{Radio follow-up observations}
In 2011, the year when Sw\,J1644+57 was discovered, very few TDEs had received radio follow-up observations (exceptions are NGC~5905, \citealt{komossa02}, and SDSS-TDE2, \citealt{vv11}). This lack of data allowed for the exciting possibility of TDE unification via orientation, akin to the successful unification model of AGN \citep{bgm+11}. The hope was that every TDE produces a relativistic jet, with Sw\,J1644+57 being the first discovered on-axis example of this new class of transient jets from SMBHs {(e.g.~\citealt{gm11})}.  
The first hint that this unified model of thermal and relativistic TDEs is not correct was presented by \citet{bower11}, who found no detections in archival radio observations available for a small number of TDEs (see Table~\ref{tab:limits}). The first targeted follow-up observations of thermal TDEs were obtained using the VLA \citep{bower13,vv13} and also yielded no detections. 
% at the time of these VLA observations it was still called the EVLA, because it was in the middle of the upgrade to what we now call the Janksy VLA. 

To obtain an estimate of the off-axis radio flux of an event similar to Sw\,J1644+57, \citet{vv13} used the observed radio light of this source and adopted a constant late-time bulk Lorentz factor of 2; this Lorentz factor was inferred from the late-time radio light curve of Sw\,J1644+57 \citep{bzp+12}. Even with this conservative assumption, the radio upper limits obtained in 2013 could be used to rule out a unified model in which all thermal TDEs produce jets similar to Sw\,J1644+57 \citep{vv13}. A few years later, a more physical prediction of the off-axis radio luminosity of jet similar to Sw\,J166+57 was presented by \citet{mgm+15}. These authors tuned a hydrodynamical simulation of a fallback-powered jet interacting with a circumnuclear medium to obtain an excellent match to the observed light curve of Sw\,J1644+57. This simulation includes the deceleration of the jet, finding that the late-time radio emission becomes isotropic after about 3 years. At this point the luminosity at 10~GHz is predicted to be $\sim 10^{40}~\rm erg\, s^{-1}$, which is larger than almost all upper limits that have been obtained. We can thus conclude that jets similar to Sw\,J1644+57 are rare; at the time of writing, 27 thermal TDEs (see \xraychap ~and \optchap) have received radio follow-up, yet none show evidence for jets as powerful as Sw\,J1644+57.

A direct comparison of radio upper limits for known TDEs to off-axis light curves based on Sw\,J1644+57 has one caveat. Beside the jet power, the radio light curve depends on the circumnuclear medium. The effect of this second parameter was studied by \citet{gmm+17}, who considered a wide range of nuclear densities (as motivated by a physical model for gas supply from stellar winds; \citealt{Generozov15}) to obtain an upper limit to the jet power for  TDEs with radio follow-up observations. For nearly all thermal TDEs studied by \citet{gmm+17}, the upper limit to the total jet energy is an order of magnitude lower than the total of Sw\,J1644+57. %{Nevertheless, less-luminous jets of comparable energy are possible -- the jet in Arp 299-B AT1 was found to have a comparable kinetic energy to the jet in Sw\,J1644+57, but was several orders of magnitude less luminous at peak due to a combination of off-axis viewing angle, higher density, and larger fraction of energy in magnetic fields \citep{mpe+18,ebz+18}.} 
{This conclusion is consistent with the work of \cite{mpe+18}, who find that the radio transient Arp 299-B AT1 is an off-axis jet with a similar energy as Sw\,J1644+57. While Arp 299-B AT1 was detected in a nearby galaxy, even at the higher redshift of most thermal TDEs ($z\sim 0.1$), the isotropic radio emission of this source would exceed most of the published upper limits (Table \ref{tab:limits}). If events like Arp 299-B AT1 are common, many more should have been detected in our current follow-up observations of thermal TDEs found in blind optical/X-ray surveys, thus confirming that powerful jets following a TDE are rare.}

In 2014, ASASSN-14li (Section~\ref{sec:14li}) was discovered and we learned that thermal TDEs can be accompanied by a low energy radio flare ($E\sim 10^{47-48}$~erg, compared to $E\sim 4\times10^{51}$~erg for Sw\,J1644+57). At the time of its discovery, the radio luminosity of ASASSN-14li was lower than any of the upper limits obtained within one year of peak (mainly thanks to the low redshift of ASASSN-14li). Furthermore, since the radio emission was observed to fade quickly, the late-time follow-up observations of previous TDEs were also not sensitive enough to catch a radio flare similar to ASASSN-14li. Most models for the origin of the radio emission from ASASSN-14li suggested this emission could be (or should be) common to all thermal TDEs (see Section~\ref{sec:14li}), which renewed the interest in rapid and sensitive radio follow-up of new TDEs. 

At the time of writing, ASASSN-14li is {one of two} optically selected TDEs (see Table~1 of \optchap\ in this volume) with published radio detections. Unlike many optical TDEs, ASASSN-14li also has bright X-ray emission, as do {at least three} of the other {five} detected radio-quiet TDEs. (Arp 299-B AT1's X-ray emission, if present, would have been masked by {the large column density towards the TDE, as the AGN in Arp 299-B is Compton thick \citep{ptak15}, while CNSS J0019+00 was not observed in the X-ray at early times.}) The first upper limits that rule out a radio flare similar to ASASSN-14li were obtained for the optical TDE iPTF-16fnl \citep{blag17}. However the low optical and X-ray luminosity of iPTF-16fnl ($L_X\sim 10^{37}~\rm erg\,s^{-1}$), plus the rapid decay rate of its optical light curve make this source an outlier in the class of thermal TDEs---although we note that these faint TDEs dominate the volumetric rate \citep{vv18b}. The radio non-detection for iPTF-16fnl could be explained by its low luminosity. More recently, \citet{vv18} presented rapid and sensitive radio follow-up observations for optical TDE PS18kh/AT2018zr \citep{Holoien19} that also rule out a radio flare similar to ASASSN-14li. The optical properties of AT2018zr are similar to ASASSN-14li, and like ASASSN-14li AT2018zr also showed a soft, thermal X-ray spectrum \citep{vv18}. However the X-ray luminosity of AT2018zr is two orders of magnitude lower than ASASSN-14li's. The recent X-ray TDE XMMSL2 J1446+68 also shows no radio emission down to deep limits, suggesting that bright X-ray emission alone is not sufficient to produce detectable radio emission {\citep{srk+19}.}

In summary, at this point we can conclude that $\sim$ year-long radio flares with a peak luminosity $\sim 10^{38}~\rm erg\,s^{-1}$ are not common to ${\it all}$ thermal TDEs. Of the {six} sources that have been studied with sufficient sensitivity, only ASASSN-14li and XMMSL1 0740-85 showed this feature. {Somewhat brighter (peak luminosity $\sim$few $10^{38}$ erg s$^{-1}$), longer-lasting radio flares similar to Arp 299-B AT1 or IGR J12580+0134 are also not ubiquitous.} If this trend is confirmed in a larger sample size, models that predict all TDEs should exhibit some level of radio emission will be disfavored (e.g. Section \ref{sec:unbound}). While as yet there is no unifying physical characteristic that distinguishes the radio-detected thermal TDEs from those without faint radio emission, ongoing observing campaigns with the world's most sensitive radio and mm telescopes are expected to greatly expand the number of TDEs with radio detections and/or deep limits, allowing us to discriminate among the models presented in Section \ref{sec:14li}.

% Svv: this topic is partially covered my the text above. We can try to quantify it in this subsection, but 
\subsection{Rate of Sw\,J1644+57-like relativistic jets}\label{sec:jetrate}
 
 While the fraction of TDEs producing faint ASASSN-14li-like radio emission is still poorly constrained, existing observations do allow us to constrain the rate of radio-loud TDEs. Based on the first 10 years of {\it Swift}/BAT monitoring, the observed rate of events similar to Sw\,J1644+57 is $\sim 10^{-2} \rm~Gpc^{-3} yr^{-1}$ \citep{mwb15}.\footnote{{assuming all TDEs that launch radio-loud jets also produce bright $\gamma$-ray emission when viewed on-axis.}} Applying a correction for the Doppler boosting of the X-rays of $f_{\rm beam} \sim 100$ \citep{vanVelzen18_ngVLA} we find $\dot{N}_{\rm J1644}\sim 1\rm~Gpc^{-3} yr^{-1}$. 
 
 To estimate the expected fraction of observed TDEs with jets similar to Sw\,J166+57 we consider the ratio of the  beaming-corrected rate to the rate of thermal TDEs, 
 \begin{equation}
 f_{\rm J1644} \equiv \dot{N}_{\rm J16644} / \dot{N}_{\rm thermal} \quad. 
 \end{equation}
 Since the volumetric rate of thermal TDEs depends on the flare's luminosity \citep{vv18b}, and most surveys are biased to the brighter events, we adopt  $\dot{N}_{\rm thermal} = 10^{-4.3}\rm ~galaxy^{-1}yr^{-1} = 5\times 10^{-5}\rm ~galaxy^{-1}yr^{-1}$. {This rate is consistent with the observational TDE rates reported for several recent optical surveys (SDSS: $3^{+5}_{-3} \times 10^{-5}\rm ~galaxy^{-1}yr^{-1}$, \citealt{vf14}; ASAS-SN: $4.1^{+12.9}_{-1.9} \times 10^{-5}\rm ~galaxy^{-1}yr^{-1}$, \citealt{hkp+16}; iPTF: ${1.7^{+2.9}_{-1.3} \times 10^{-4}\rm ~galaxy^{-1}yr^{-1}}$, \citealt{hgc+18}; see also the detailed discussion by \optchap~in this volume).} The density of galaxies that can produce detectable TDEs is $\sim 10^{-2.4}~\rm Mpc^{-3}$ \citep[for $h=0.7$; see][]{vf14}, and we thus obtain $\dot{N}_{\rm thermal}\sim 10^{2.3} ~\rm Gpc^{-3}yr^{-1}$. 
 
These two estimates of the rate imply $f_{\rm J1644}\sim 10^{-2.3} (f_{\rm beam}/100)$. One caveat to this estimate relates to the redshift evolution of the disruption rate; the jetted TDEs are found at higher redshift ($z\sim 1$) compared to the thermal TDEs ($z\sim 0.1$) and the redshift evolution of the TDE rate is unknown (depending on the loss-cone feeding mechanism the rate could both increase or decrease with redshift, see \ratechap\ in this volume). At the time of writing, at least {29} TDEs have received radio follow-up observations that rule out a jet similar to Sw\,J1644+57 (Table \ref{tab:limits}, plus the {six} TDEs with faint radio detections). This lack of bright sources is consistent with our estimate of $f_{\rm J1664}$. Unless we underestimated the beaming factor, we should expect powerful jets in at most 1\% of thermal TDEs. Constraints on the beaming factor can be obtained by measuring the isotropic rate of events similar to Sw\,J1644+57 using blind radio surveys. 

% this is covered in the final paragraph of the upper limit section, don't think we can quantify this more 
% \subsubsection{Rate of Sw J1644-like relativistic jets}

\subsection{Blind Searches}\label{sec:blind}

Radio emission from TDEs has so far been detected {primarily} via follow-up observations (as TDEs have been discovered at other wavelengths, e.g., X-ray, UV, and optical). {The sample of radio TDEs thus obtained is inevitably observationally biased, as the subset of TDEs chosen for radio follow up to date has not been uniformly selected.} Direct discovery of TDEs in the radio can be achieved by blind searches in radio data. Several such searches have been conducted either by using archival data or by carrying out specifically designed time-domain radio surveys (e.g., \citealt{fkh+94, cif03, bsb+07, cbk+11, ofb+11, fko+12, hbw+13, mhb+16})\footnote{A more complete list of radio transient surveys and additional details can be found at  http://www.tauceti.caltech.edu/kunal/radio-transient-surveys/index.html.}. A blind search in the radio is more challenging compared, for example, to optical surveys, as the field of view of radio interferometers is much narrower and the typical sensitivity of radio facilities enables detection of radio emission from thermal TDEs only from nearby events. With time, the capabilities of numerous facilities have been upgraded, thus enabling wider and more sensitive blind searches. 

%The results of past radio blind searches can be used to derive an upper limit on the rate of TDEs. 
Compared with theoretical expectations, most of the blind searches have not been wide and/or deep enough to discover TDEs. However, the upgrade of the VLA over the last decade has opened a window for performing effective blind surveys in which TDEs can be expected to be discovered. One such survey is the the Caltech-NRAO Stripe 82 Survey (CNSS; \citealt{mhb+16}). This survey was designed to cover the full $270\deg^2$ of the SDSS Stripe 82 area over multiple timescales in the $2-4$ GHz band {with high spatial resolution ($\sim3$") and achieved a per-epoch sensitivity of $\sim80$ $\mu$Jy \citep{mhb+16}. In five epochs spread over two years, one TDE was discovered, CNSS J0019+00 \citep{amh+19}. This implies that the rate of TDEs with radio emission brighter than 0.5 mJy is $1.8^{+5.4}_{-1.6}\times10^{-3}$ deg$^{-2}$ \citep{amh+19}. This corresponds to a volumetric rate of $\sim10$ Gpc$^{-3}$ y$r^{-1}$ for radio TDEs with the same peak luminosity as CNSS J0019+00.} 

{At face value, comparing this rate to the results presented above implies that TDEs with low-luminosity radio outflows similar to CNSS J0019+00 or ASASSN-14li are $\sim10$ times more common than Sw\,J1644+57-like radio-loud TDEs, and comprise $\sim10$\% of thermal TDEs \citep{amh+19}. However, a direct comparison} is not trivial due to remaining uncertainties in the TDE occurrence rate {and its redshift evolution, the fraction of TDEs missed by optical/X-ray surveys}, and the radio luminosity function of thermal TDEs. Additionally, identifying off-axis jetted TDEs as transient sources may be challenging because their radio emission evolves very slowly; during the 1.5 yr of the CNSS pilot survey, the 3 GHz flux density of a Sw\,J1644+57-like off-axis jet may not have changed sufficiently to be flagged as a variable source \citep{ebz+18}. Radio surveys with more widely-separated epochs may be better suited to detect slow synchrotron transients like off-axis jetted TDEs \citep{mwb15}.

{On the other hand, if} most TDEs are thermal events that produce mostly weak radio emission, {as implied by by the rate from the CNSS,} then a wider survey with the same sensitivity as the CNSS may be {more effective to} blindly detect TDEs. In fact, such a survey, the VLA All Sky Survey (VLASS), has begun at the time of writing this book. The VLASS is a is an all sky survey that will be carried out using the VLA over the next several years. VLASS is using a drift scan technique that allows fast mapping of the sky \citep{mmf+19}. Each piece of survey sky is expected to be observed three times during the project with each epoch separated by $\sim 18-24$ months. According to current estimates, tens of TDEs will be discovered in VLASS (\citealt{mwb15,vanVelzen18_ngVLA}; {\citealt{amh+19}}). However, this estimate is highly dependent on the true TDE rates and the TDE luminosity function, as already discussed above.

{In the southern hemisphere, the best current facilities for wide-field radio transient searches are the Australian Square-Kilometre-Array Pathfinder (ASKAP;  \citealt{askap}) and MeerKAT \citep{meerkat,css+18}. ASKAP utilizes a new phased array feed receiver design that effectively gives it an instantaneous field of view much larger than the VLA's ($\sim30$ deg$^2$ versus 0.25 deg$^2$), making it well-suited for targeted follow up of poorly-localized transients (including gravitational wave events, as recently demonstrated by \citealt{dsm+19}) or for wide-field blind transient detection surveys. Several such surveys are already planned or underway \citep{mck+13,hbm+16,hhb+16,bbm+18}. The largest of these is VAST (the ASKAP survey for Variables and Slow Transients), which includes a wide-field component (VAST-Wide) that will survey 10,000 deg$^2$ day$^{-1}$ over a two-year period with an RMS sensitivity of 0.5 mJy. VAST-Wide is predicted to discover up to a dozen TDEs per year \citep{vkf11,mck+13}. MeerKAT, located in South Africa, has a smaller but still impressive field of view ($\sim 1^{\circ}$ at 1.4 GHz; \citealt{dmb+20}) and has achieved sub-microjansky noise in its deepest observations to date \citep{mcc+20}. Transient science has been designated as one of the key early science priorities for MeerKAT under the ThunderKAT Large Survey Project \citep{fwa+17}, which recently published its first blind detection of a radio transient \citep{dmb+20}.} %ThunderKAT includes weekly monitoring of key fields, target-of-opportunity follow up of transients discovered at other wavelengths, and commensal transient searches carried out in data obtained for the other MeerKAT large survey projects. It is expected to detect a few tens of TDEs over its initial 5-year duration \citep{fwa+17}.} 

{Both ASKAP and MeerKAT are currently limited to observing frequencies $\leq1.4$ GHz, and therefore using them to blindly discover TDEs involves many of the same challenges as discovering TDEs in the CNSS and VLASS (including the slow evolution of radio TDEs at low frequencies, and the challenge of distinguishing TDEs from background variable AGN). However, both ASKAP and MeerKAT are precursors to the Square Kilometer Array (SKA; \citealt{ska}), which is expected to include high-frequency sensitivity up to at least 15 GHz.\footnote{ {https://astronomers.skatelescope.org}}} Radio observations with future facilities such as the SKA and the next generation VLA (ngVLA)\footnote{{While the ngVLA has a much smaller field of view than the SKA under the current design, its exquisite sensitivity at high frequencies will make it a key tool for the follow up and discovery of TDEs. Additionally, its milliarcsecond resolution will be very interesting for VLBI observations of nearby events \citep{vanVelzen18_ngVLA}.}} are expected to produce larger samples of TDEs {(hundreds per year)} that are also more distant at higher redshifts \citep{mwb15,drf+15,vanVelzen18_ngVLA}.

\section{Conclusions}

Radio observations of TDEs provide unique information about these rare events that cannot be gleaned from observations at other wavelengths. In particular, they probe fast-moving material ejected from the system, in contrast to the generally thermal emission seen in the optical, UV, and X-rays. Radio observations have shown that a small subset of TDEs produce relativistic jets, while a larger fraction may produce less energetic outflows. The physical conditions required for jet production and the diversity of sub-parsec scale nuclear gas profiles around SHBMs in TDE hosts will be revealed by increasing the sample size of radio-detected TDEs. {Given the broad range in observed TDE radio luminosities, we suggest that radio follow up of new events should optimally probe radio luminosities $\nu F_{\nu} \leq 10^{37}$ erg s$^{-1}$, as many existing upper limits are too shallow to detect outflows similar to those observed in e.g.~ASASSN-14li. Furthermore,} with the advent of ALMA, it is now possible to probe even less energetic outflows, which are expected to reach higher peak luminosities in the millimeter than in the centimeter bands probed by the VLA \citep{ywl+16}. 

The increased TDE discovery rate promised by new sensitive, wide-field optical surveys like ZTF is already being realized \citep{vv18}. Simultaneously, new and upcoming radio surveys may soon provide the first population of TDEs discovered in the radio band ({\citealt{mck+13}}, \citealt{mwb15}, {\citealt{fwa+17,amh+19}}). Within the next five years, we will have placed strong constraints on the rate of weak radio outflows in TDEs, and may discover further extremes in an already diverse radio TDE population. The sample of radio TDEs will continue to grow in the era of LSST plus SKA \citep{drf+15} --- although we note that the slow evolution of the radio light curve in {low-frequency} observations ($1.4$~GHz) will present a major challenge for TDE identification \citep{dr15}. A time-domain survey at higher frequencies ($\gtrsim 5$~GHz) with the ngVLA \citep{vanVelzen18_ngVLA} {or SKA band 5} would be the ultimate discovery machine for radio-emitting TDEs, finding as many as $\sim 10^2$ jetted TDEs per year.

\begin{center}
\begin{table}
\caption{Published Radio Upper Limits of TDEs.}
\label{tab:limits} 
\begin{tabular}{clcccccl}
%\hline \noalign{}
\hline\noalign{\smallskip}
{Fig.~\ref{fig:radio}} & Name & $z$ & MJD & $\Delta t^1$  & Frequency & $3\sigma$ upper & Reference \\
{label} & & & (d) & (yr)     &    (GHz) & limit ($\mu$Jy) {\smallskip} & \\
\hline\noalign{\smallskip}
{1} & RXJ1624+7554 & 0.06 & 56082.342 & 21.6 & 3.0 & 51 & \cite{bower13}\\
\hline\noalign{\smallskip}
{2} & RXJ1242-1119 & 0.046 & 56085.027 & 19.9 & 3.0 & 54 & \cite{bower13}\\
\hline\noalign{\smallskip}
{3} & SDSSJ1323+48 & 0.08 & 56082.301 & 8.6 & 3.0 & 102 & \cite{bower13}\\
\hline\noalign{\smallskip}
{4} & SDSSJ1311-01 & 0.156 & 56085.040 & 8.3 & 3.0 & 57 & \cite{bower13}\\
\hline\noalign{\smallskip}
{5} & NGC5905 & 0.012 & 50390.786 & 6.3 & 8.46 & 90 & \cite{komossa02}\\ 
{5} & NGC5905 & 0.012 & 56082.328 & 21.9 & 3.0 & 200 & \cite{bower13}\\
\hline\noalign{\smallskip}
{6} & GALEX-D1-9 & 0.326 & 55955.988 & 7.46 & 5.0 & 27 & \cite{vv13}\\
\hline\noalign{\smallskip}
%GALEX-D3-13 & 0.3698 & 52975.740 & -0.113 & 1.4 & 150 & \cite{bower11}\\
%GALEX-D3-13 & 0.3698 & 53478.312 & 1.26 & 1.4 & 150 & \cite{bower11}\\
%GALEX-D3-13 & 0.3698 & 53516 & 1.37 & 1.4 & 150 & \cite{bower11}\\ % date quoted in Bower 2011 is wrong; the VLA did not observe D3-13 on that date.
% Bower 2011ApJ...732L..12B: observations from  2003 December 2, 2005 April 18, and 2005 June 20. Individual rms <50 muJy
{7} & GALEX-D3-13$^2$ & 0.3698 & 53478.312 & 1.26 & 1.4 & 45 & \cite{bower11}\\ % 15 muJy rms based on co-add from co-add of 3 obs listed above
{7} & GALEX-D3-13 & 0.3698 & 55955.498 & 8.05 & 5.0 & 24 & \cite{vv13}\\
\hline\noalign{\smallskip}
{8} & D23H-1 & 0.186 & 55955.938 & 4.8 & 4.3 & 24 & \cite{vv13}\\
\hline\noalign{\smallskip}
{9} & PTF10iya & 0.224 & 55955.479 & 1.6 & 5.0 & 24 & \cite{vv13}\\
\hline\noalign{\smallskip}
{10} & PS1-10jh & 0.170 & 55649.348 & 0.71 & 5.0 & 45 & \cite{vv13}\\
\hline\noalign{\smallskip}
{11} & SDSS-TDE1 & 0.136 & 55955.942 & 5.4 & 5.0 & 30 & \cite{vv13}\\
\hline\noalign{\smallskip} % 
{12} & SDSS-TDE2 & 0.252 & 54407.071 & 0.137 & 8.4 & 255 & \cite{vv11} \\ % rms=85 muJy, from co-add of observations 7&92 days after first detection; used midpoint to compute delta T
{12} & SDSS-TDE2 & 0.252 & 55955.935 & 4.6 & 5.0 & 36 & \cite{vv13}\\
\hline\noalign{\smallskip}
{13} & SDSSJ1201+30 & 0.146 & 55819.767 & 1.267 & 8.3 & 108 & \cite{saxton12} \\  
{13} & SDSSJ1201+30 & 0.146 & 55819.779 & 1.267 & 4.8 & 135 & \cite{saxton12} \\ 
{13} & SDSSJ1201+30 & 0.146 & 55819.788 & 1.267 & 1.4 & 201 & \cite{saxton12} \\
\hline\noalign{\smallskip}
{14} & PTF09axc & 0.1146 & 56836.221 & 4.953 & 6.1 & 150 & \cite{arc14}\\
{14} & PTF09axc & 0.1146 & 56836.233 & 4.953 & 3.5 & 330 & \cite{arc14}\\
\hline\noalign{\smallskip}
{15} & PS1-11af & 0.4046 & 55649.053 & 0.186 & 4.9 & 51 & \cite{chor14}\\
{15} & PS1-11af & 0.4046 & 55933.409 & 0.965 & 5.875 & 30 & \cite{chor14}\\
{15} & PS1-11af & 0.4046 & 56443.992 & 2.363 & 5.875 & 45 & \cite{chor14}\\
\hline\noalign{\smallskip}
{16} & PS16dtm  & 0.0804 & 57653.227 & 0.00610 & 21.8 & 57 & \cite{blan17} \\
{16} & PS16dtm  & 0.0804 & 57653.250 & 0.00616 & 6.0 & 23 & \cite{blan17} \\
{16} & PS16dtm  & 0.0804 & 57743.136 & 0.252 & 21.8 & 51 & \cite{blan17} \\
{16} & PS16dtm  & 0.0804 & 57743.159 & 0.252 & 6.0 & 25 & \cite{blan17} \\
\hline\noalign{\smallskip}
{17} & iPTF16fnl & 0.016328 & 57631.281 & -0.00224 & 22 & $52^3$ & \cite{blag17} \\
{17} & iPTF16fnl & 0.016328 & 57631.293 & -0.00221 & 6.1 & $34^3$ & \cite{blag17} \\
{17} & iPTF16fnl & 0.016328 & 57632 & -0.000274 & 15 & 117 & \cite{blag17} \\
{17} & iPTF16fnl & 0.016328 & 57636 & 0.0107 & 15 & 117 & \cite{blag17} \\
{17} & iPTF16fnl & 0.016328 & 57648 & 0.0435 & 15 & 117 & \cite{blag17} \\
{17} & iPTF16fnl & 0.016328 & 57683 & 0.139 & 15 & 75 & \cite{blag17} \\
\hline\noalign{\smallskip}
{18} & iPTF15af & 0.07897 & 57053.215 & -0.0659 & 22 & 108 & \cite{blag19} \\
{18} & iPTF15af & 0.07897 & 57053.225 & -0.0659 & 6.1 & 84 & \cite{blag19} \\
\hline\noalign{\smallskip}
{19} & AT2018zr & 0.071 & 58205.883 & 0.0344 & 16 & 120 & \cite{vv18} \\
{19} & AT2018zr & 0.071 & 58207.170 & 0.0379 & 10 & 27 & \cite{vv18} \\
{19} & AT2018zr & 0.071 & 58236.154 & 0.117 & 10 & 37.5 & \cite{vv18} \\
\hline\noalign{\smallskip}
{20} & AT\,2018fyk & 0.059 & 58380.677 & 0.0313 & 18.95 & 38 & \cite{wpv+19} \\
{20} & AT\,2018fyk & 0.059 & 58407.490 & 0.1048 & 18.95 & 74 & \cite{wpv+19} \\
{20} & AT\,2018fyk & 0.059 & 58444.289 & 0.2055 & 18.95 & 53 & \cite{wpv+19} \\
\hline\noalign{\smallskip}
{21} & AT\,2017eqx & 0.1089 & 57948.465 & 0.0736 & 21.7 & 80 & \cite{nbb+19} \\
{21} & AT\,2017eqx & 0.1089 & 57948.477 & 0.0736 & 6.0 & 27 & \cite{nbb+19} \\
{21} & AT\,2017eqx & 0.1089 & 57987.240 & 0.180 & 21.7 & 76 & \cite{nbb+19} \\
{21} & AT\,2017eqx & 0.1089 & 57987.252 & 0.180 & 6.0 & 26 & \cite{nbb+19} \\
\hline\noalign{\smallskip}
{22} & XMMSL2\_J1446+68 & 0.029 & 57646.155 & 0.0643 & 21.7 & 85 & {\cite{srk+19}}	\\
{22} & XMMSL2\_J1446+68 & 0.029 & 57646.167 & 0.0643 & 6.0 & 28 & {\cite{srk+19}}	\\
{22} & XMMSL2\_J1446+68 & 0.029 & 57806.511 & 0.503 & 21.7 & 84 & {\cite{srk+19}} \\
{22} & XMMSL2\_J1446+68 & 0.029 & 57806.524 & 0.503 & 6.0 & 18 & {\cite{srk+19}} \\
\hline\noalign{\smallskip}
{23} & {ASASSN-18pg} & {0.017392} & {58319.454} & {-0.06611} & {18.95} & {50} & {\cite{hat+20}} \\
{23} & {ASASSN-18pg} & {0.017392} & {58336.617} & {-0.01912} & {18.95} & {43} & {\cite{hat+20}} \\
\hline\noalign{\smallskip}
\end{tabular}
\text{1. Time relative to peak bolometric luminosity, given in the observer frame. For events with sparse multi-wavelength coverage,} 
\text{ we list the time relative to maximum light in a single optical or X-ray observing band.}
\text{2. Based on serendipitous observations of the TDE field co-added by \cite{bower11}. (Note: \citealt{bower11} reports the observation}
\text{ dates as 2003 Dec 2, 2005 Apr 18, and 2005 Jun 20. However, the VLA data archive records no observations within 1 degree}
\text{ of the TDE coordinates on Jun 20, so it is likely that one or more of the observations of the field collected under the same}
\text{ observing program between 2005 May 14 - 2005 Jun 7 were used instead. We use the MJD of the 2005 Apr 18 observation to}
\text{ compute $\Delta t$ for this observation.)}\\
\text{3. Based on a re-reduction of the data by KDA.}
\end{table}
\end{center}

\begin{acknowledgements} 

We acknowledge useful discussions with the attendees of the ISSI TDE workshop in October 2018, in particular Tsvi Piran. KDA acknowledges
support provided by NASA through the NASA Hubble Fellowship grant HST-HF2-51403.001 awarded by the Space Telescope Science Institute, which is operated by the Association of Universities for Research in Astronomy, Inc., for NASA, under contract NAS5-26555. BAZ acknowledges support while serving at the National Science Foundation (NSF) and from the Dark Cosmology Centre (DARK) at the University of Copenhagen.  Any opinion, findings, and conclusions expressed in this material are those of the authors and do not necessarily reflect the views of the supporting agencies.
\end{acknowledgements}

\bibliographystyle{aps-nameyear}
\bibliography{ms}

\end{document}